\documentclass{ws-ijmpcs_2}
\usepackage{amssymb}
\usepackage{float}
\usepackage[dvipsnames]{xcolor}

\begin{document}

\markboth{G. A. Contrera et. al.}
{Hybrid stars in the framework of different NJL models}

%
\catchline{}{}{}{}{}
%

\title{Hybrid Stars in the Framework of different NJL Models}

\author{G. A. Contrera$^{a,b,c,\dag}$,
M. Orsaria$^{a,b,d,\ddag}$,
I.~F.~Ranea-Sandoval$^{a,b,\S}$ and
F. Weber$^{d,e,\P}$
\vspace{6px}}

\address{
$^{a}$ Grupo de Gravitaci\'on, Astrof\'isica y Cosmolog\'ia,\\
  Facultad de Ciencias Astron{\'o}micas y Geof{\'i}sicas, Universidad Nacional de La Plata,\\
  Paseo del Bosque S/N (1900), La Plata, Argentina.\\
$^{b}$ CONICET, Godoy Cruz 2290, 1425 Buenos Aires, Argentina.\\
$^{c}$ IFLP, UNLP, CONICET, Facultad de Ciencias Exactas, calle 49 y 115, La Plata, Argentina.\\
$^{d}$ Department of Physics, San Diego State University, 5500 Campanile Drive,\\
  San Diego, CA 92182, USA.\\
$^{e}$ Center for Astrophysics and Space Sciences, University of California,\\
  San Diego, La Jolla, CA 92093, USA.\\
$^{\dag}$contrera@fisica.unlp.edu.ar, $^{\ddag}$morsaria@fcaglp.edu.ar,\\
$^{\S}$iranea@fcaglp.unlp.edu.ar, $^{\P}$fweber@mail.sdsu.edu
}

\maketitle

\begin{abstract}
We compute models for the equation of state (EoS) of the matter in the cores
of hybrid stars. Hadronic matter is treated in the non-linear
relativistic mean-field approximation, and quark matter is modeled by
three-flavor local and non-local Nambu$-$Jona-Lasinio (NJL) models
with repulsive vector interactions. The transition from hadronic to
quark matter is constructed by considering either a soft phase
transition (Gibbs construction) or a sharp phase transition (Maxwell
construction). We find that high-mass neutron stars with masses up to
$2.1-2.4 M_\odot$ may contain a mixed phase with hadrons and quarks in
their cores, if global charge conservation is imposed via the Gibbs
conditions. However, if the Maxwell conditions is considered, the
appearance of a pure quark matter core either destabilizes the star
immediately (commonly for non-local NJL models) or leads to a very
short hybrid star branch in the mass-radius relation (generally for
local NJL models).

\end{abstract}

\ccode{PACS numbers:26.60.-c, 26.60.Kp, 25.75.Nq, 97.60.Jd}

\section{Introduction}

The question of whether or not quark matter exists in neutron stars
has received renewed interest (see
Refs.\ \refcite{Orsaria:2012je,Orsaria:2013hna,Weber:2014qoa}, and
references therein) by the discovery of the two massive neutron stars
(NS) J1614--2230 ($1.97\pm0.04M_{\odot}$\cite{Demorest:2010bx}, recently
updated to $1.928\pm0.017M_{\odot}$\cite{Fonseca:2016tux}) and
J0348+0432 ($2.01\pm0.04M_{\odot}$\cite{Antoniadis13}). Aside from
these two massive objects, there may exist even heavier neutron stars
which are known as "Black Widow Pulsars", such as B1957+20
($2.39^{+0.36}_{-0.29}M_{\odot}$\cite{vanKerkwijk:2010mt}).  If the
dense interior of a NS is indeed converted to quark
matter\cite{glendenning00,weber99:book,weber05:a}, it must be
three-flavor quark matter since it has lower energy than two-flavor
quark matter. And just as for the hyperon content of NS, strangeness
is not conserved on macroscopic time scales, which allows them to
convert confined hadronic matter to three-flavor quark matter until
equilibrium brings this process to a halt. As first realized by
Glendenning \cite{glendenning00}, the presence of quark matter in
these stars enables the hadronic regions of the mixed phase to become
more isospin symmetric than in the pure phase by transferring electric
charge to the quark phase. The symmetry energy can be lowered thereby
at only a small cost in rearranging the quark Fermi surfaces. The
stellar implication of this charge rearrangement is that the mixed
phase region of a NS will have positively charged regions of nuclear
matter and negatively charged regions of quark matter
\cite{glendenning00}. This should have important implications for its
electric and thermal properties.

It has been shown \cite{Alford:2001zr,Voskresensky:2002hu,Tatsumi:2002dq} that the appearance of a mixed phase of quarks and
hadrons in NS is favored when the value of the surface tension between
nuclear matter and quark matter is lower than $\sigma \sim 40$ MeV/fm$^2$
.
Some recent studies \cite{Palhares:2010be,Endo:2011em,Pinto:2012aq,Ke:2013wga,Mintz:2012mz,Yasutake:2014oxa} predict the value of the surface tension around $\sigma \sim 5-30$ MeV/fm$^2$, while others \cite{Alford:2001zr,Voskresensky:2002hu,Tatsumi:2002dq,Carmo:2013fr,Lugones:2013ema} obtain a value around $\sigma \sim 50-300$ MeV/fm$^2$. Comparing those works it can be seen that the results are strongly model dependent. Thus, the appearance of a mixed quark-hadron phase in NS is therefore an open issue.

Our study is based on NS containing deconfined quark matter,
i.e.\ quark-hybrid stars (QHSs). As in previous works
\cite{Orsaria:2012je,Orsaria:2013hna,Ranea-Sandoval:2015ldr,Spinella:2015ksa,%
  deCarvalho:2015lpa}, we use three-flavor local and non-local
Nambu$-$Jona-Lasinio (NJL) models with repulsive vector interactions
to describe the quark matter phase. To model hadronic matter, we adopt
the non-linear Walecka model using different nuclear parametrizations
adjusted to the properties of infinite nuclear matter at saturation
density. A phase transition between hadronic matter and quark matter
is constructed via the Gibbs and Maxwell conditions, depending on the
assumption of the surface tension at the hadron-quark interface.

\section{Hadron-Quark phase transition}
\label{sec:trans}

Several theoretical works
\cite{Benic:2014jia,Buballa:2003et,Negreiros:2010tf,Kurkela:2009gj} have
shown that a first-order phase transition between hadronic and quark
matter ought to happen for cold and dense hadronic matter as existing
in the cores of NS. The density at which such a phase transition
occurs is not well defined, but it is supposed to be several times the
nuclear saturation density. To model this phase transition, two
different treatments are generally adopted. These are A) the Maxwell
construction, in which a sharp phase transition between hadronic and
quark matter takes place, which rules out the existence of a mixed
phase; B) the Gibbs construction, where pressure in the mixed phase
varies, which gives rise to the appearance of a mixed phase.

The value of the surface tension at the interface separating hadrons
from quarks is crucial since it determines which scenario ought to be
used to study the phase transition. Theoretical studies \cite{Alford:2001zr,Voskresensky:2002hu,Tatsumi:2002dq} suggest that
if the surface tension between hadronic and quark matter is around
$5-40$ MeV/fm$^2$, the Gibbs construction would be favored. Otherwise
the transition should be described by the Maxwell
construction. Given the uncertainties in the value of the surface
tension, both scenarios should be considered plausible.

\begin{figure}[htb]
\centering
\includegraphics[angle=90, width=0.8 \textwidth]{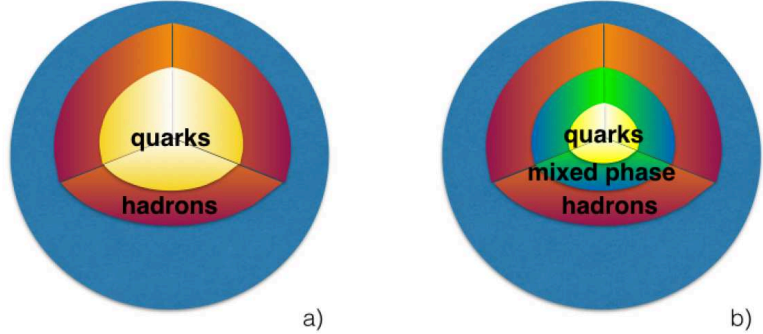}
\caption{(Color online) Schematic illustration of the interior
  structure of a hybrid star for the Maxwell construction (a) and the
  Gibbs construction (b). Regions of quark-hadron matter are only
  obtained for the latter.}
\label{fig:bol}
\end{figure}

\begin{figure}[htb]
\centering
\includegraphics[width=1.0 \textwidth]{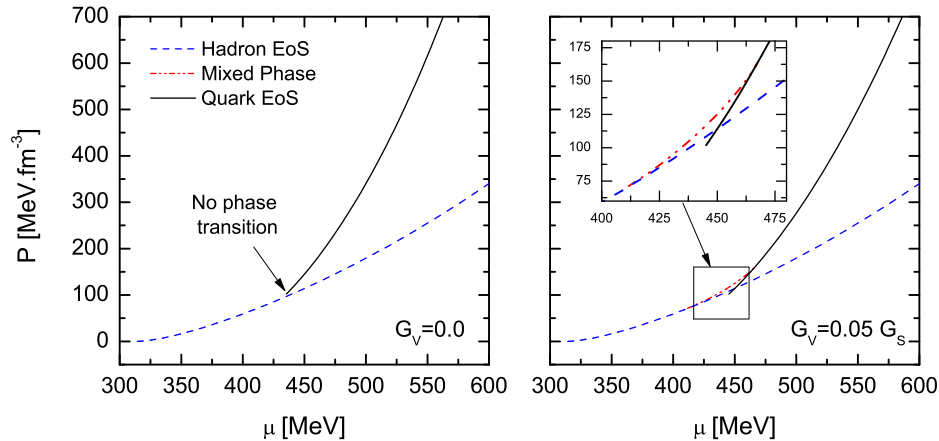}
\caption{(Color online) Pressure, $P$, as a function of quark chemical
  potential, $\mu_q$. In the left panel (without vector interactions, $G_V$=0.0), there is no hadron-quark
  phase transition since the hadronic and quark matter EoS do not
  cross.  In the panel on the right (with vector interactions, $G_V$=0.05$G_S$), the crossing of both EoS
  determines the density at which the hadron-quark phase transition
  occurs for the Maxwell construction. The insert shows the smooth
  mixed-phase region characteristic of the Gibbs transition. These plots
  correspond to the NL3 parametrization ($\chi_\sigma$=0.7) for the hadron EoS
  and non-local NJL (Set II in Ref. 22) for the quark EoS.
   }
\label{fig:comp_gibbs_max}
\end{figure}

For example, the study of the constant speed of sound (CSS)
parametrization for quark matter\cite{Alford:2013aca} assumes that the
Maxwell construction is the proper one to describe the phase
transition, which is determined by the crossing point between the
hadronic and quark matter equation of state
(Fig.\ \ref{fig:comp_gibbs_max}). This phase transition is isobaric
and occurs over a finite density range defined by
\begin{eqnarray}
P^H (\mu_B^H,\mu_e^H)&=&P^q(\mu^q,\mu_e^q) \, , \qquad
\mu_B^H = 3 \mu^q \, ,
\end{eqnarray}
where $\mu_B^H$ and $\mu^q$ are the hadronic and quark chemical
potentials, respectively. For the Maxwell transition, the baryonic
chemical potential is continuous while the electron chemical
potential, $\mu_e$, jumps at the interface between the hadronic and
quark phases. The phase transition is abrupt and the pressure within
the transition region is constant.  As a consequence, the mixed phase
region of the Maxwell transition is strictly excluded from the cores
of neutron stars, leading to a density discontinuity at the interface
between confined hadronic matter and deconfined quark matter. The
situation is drastically different for the Gibbs case,
as illustrated in Fig.\ \ref{fig:bol},
where extended regions of mixed quark-hadron matter in the cores of
neutron stars can be obtained, $\mu_e$ is continuous, and the pressure
varies monotonically with density (see
Fig.\ \ref{fig:comp_gibbs_max}). Electric charge neutrality is
fulfilled locally for the Maxwell construction, in contrast to the Gibbs case
where charge neutrality is imposed as a global constraints.  The
condition for phase equilibrium between hadronic and quark matter for the Gibbs construction is given by
\begin{eqnarray}
P^H (\mu_B^H,\mu_e^H)&=& P^q(\mu^q,\mu_e^q) \, , \qquad \mu_B^H = 3
\mu_B^q \, , \qquad \mu_e^H = \mu_e^q \, ,
\end{eqnarray}
with the baryon and electron chemical potentials continuous at the
phase boundary.

For the Maxwell construction, the necessary and sufficient condition
for the phase transition to occur can be readily seen in the pressure
versus baryon chemical potential diagram. Namely,  if there is a point at
which the equations of state of the two phases intersect, then a phase
transition occurs. This is shown in the right panel of
Fig.\ \ref{fig:comp_gibbs_max}.

For the Gibbs construction, the crossing of the quark and hadronic EoS
is a necessary condition, but it is not a sufficient one. This is so
because the Gibbs condition is satisfied by imposing global charge
conservation so that pressure, energy density, and baryon and electron
chemical potentials vary monotonically as the phase transition
proceeds. For this reason it is convenient to include a new parameter
which takes into account the volume proportion of quark matter and the
variation of the thermodynamic quantities point by point in the mixed
phase (see Refs.\ \refcite{Orsaria:2012je,Orsaria:2013hna} for details).

Since there is no sharp phase transition for the Gibbs construction
between hadronic and quark matter but rather a mixed phase, it is not
possible to perform the CSS parametrization for the quark matter EoS,
as proposed in Ref.\ \refcite{Ranea-Sandoval:2015ldr}. We also note
that, for the Gibbs treatment, stellar cores made of pure quark matter
are not obtained for the different EoS considered in this work.
However, a mixed phase of quarks and hadrons is always present,
consisting of $\sim 30-60 \%$ of quark matter for the local NJL
parametrization, and $\sim 30-35\%$ for the non-local NJL models,
depending on the combination of the hadronic and quark matter
parameters.

\begin{figure}[htb]
\centering
\includegraphics[width=0.9 \textwidth]{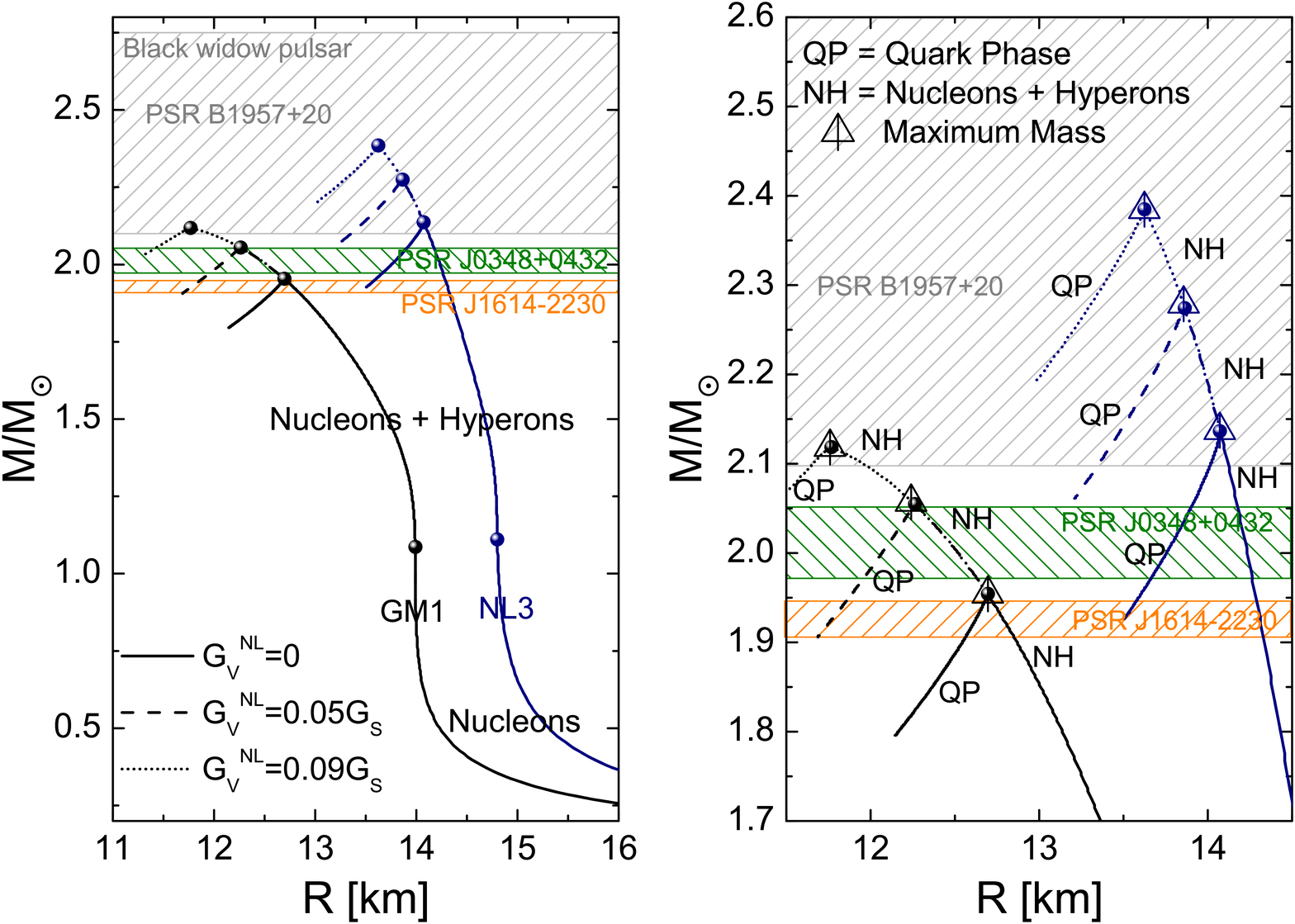}
\caption{(Color online) Mass-radius relationship of quark-hybrid stars
  for the Maxwell construction, computed for universal coupling of
  hyperons and nuclear parametrizations GM1 and
  NL3.}
\label{mr_max}
\end{figure}

\begin{figure}[htb]
\centering
\includegraphics[width=0.9 \textwidth]{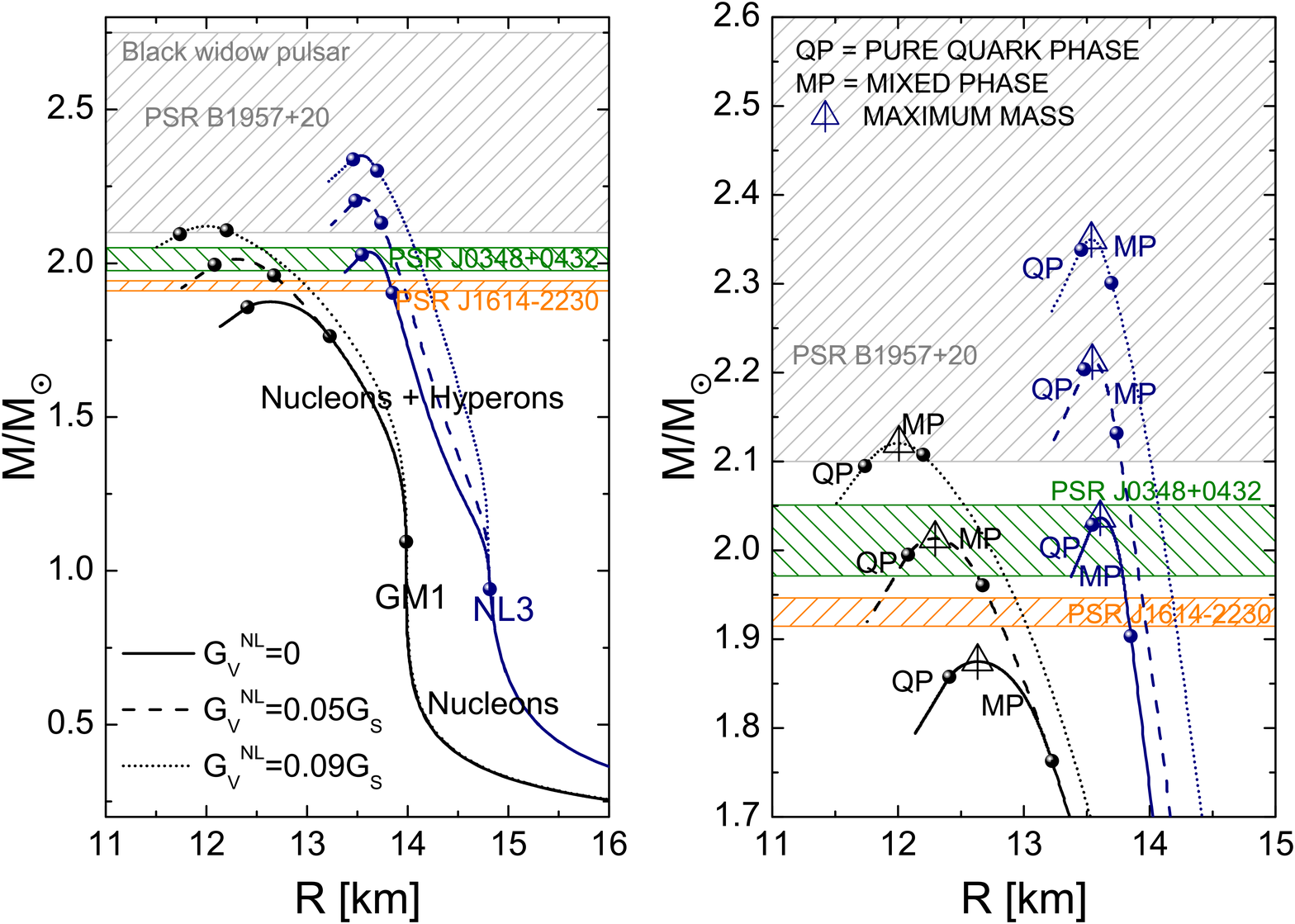}
\caption{(Color online) Same as Fig.\ \ref{mr_max} but for the Gibbs
  construction.}
\label{mr_gibbs}
\end{figure}

\section{Results and Conclusions}

The indication of this work is that if the Gibbs condition is used to
model phase equilibrium between hadronic and quark matter, high-mass
neutron stars, such as the pulsars J3048+0432,
J1614-2230 and B1957+20, may contain significant amounts of quark-hadron matter in
their cores. The established values are $30-35 \%$ of quark matter in
case of a non-local NJL model, and $\sim 30-60 \%$ for the local NJL
model, provided the surface tension between hadronic and quark matter
is low, as suggested in the recent literature
(Refs.\ \refcite{Palhares:2010be,Pinto:2012aq,Ke:2013wga,Mintz:2012mz,Yasutake:2014oxa})

For the non-local NJL model and the NL3 parametrization for
the hadronic phase, we find that pure quark matter would not exist in
stable neutron stars, since only neutron stars that lie on the left of
the mass peak have dense enough cores to contain quark matter. Such
stars, however, are unstable against radial oscillations and thus
could not exist stably in the universe. With increasing stellar mass
(that is, density) the stellar cores are composed of either nucleons,
nucleons and hyperons, or a mixed phase of nucleons, hyperons, and
quark matter (see Fig.\ \ref{mr_gibbs}).

On the other hand, the main conclusion which follows from using the
Maxwell construction is that the non-local NJL models that we have
studied generally do not lead to hybrid stars, while the local NJL
models do for some parametrizations. These stars, however, cover only
a very small range of masses and radii. This is different when the
Gibbs condition is used to determine the phase transition.  Most of the
non-local NJL EoSs lack hybrid branches because the jump in energy
density at the phase transition is so large that the quark matter core
destabilizes the star immediately (see Fig.\ \ref{mr_max}).

In the future, it would be interesting to study the impact of color
superconductivity in the quark phase, and different hadronic lagrangians (e.g.,  density dependent nuclear field theory) on the
bulk properties of neutron stars.

\bigskip
\section*{Acknowledgments}
G.A.C., M.O. and I.F.R-S thank CONICET and UNLP (Argentina) for
financial support. M.O. thanks the American Physical Society for an
International Research Travel Award. F.W. is supported by the National
Science Foundation (USA) under Grant PHY-1411708.

\end{document}